\documentclass[journal,twoside]{IEEEtran}
\IEEEoverridecommandlockouts
\usepackage{cite}
\usepackage{amsmath,amssymb,amsfonts}
\usepackage{graphicx,graphics}
\usepackage[caption=false]{subfig}
\usepackage{algorithmic}
\usepackage{textcomp}
\usepackage{tabulary}
\usepackage{array}
\usepackage{siunitx}
\usepackage{svg}
\usepackage{enumitem,kantlipsum}
\usepackage{multirow}
\usepackage{makecell}
\usepackage{pifont}
\usepackage{dblfloatfix}
\usepackage{etoolbox}
\usepackage{mathtools}
\usepackage{scalerel}
\usepackage[some,bottom]{background}

\newcolumntype{L}[1]{>{\raggedright\let\newline\\\arraybackslash\hspace{0pt}}m{#1}}
\newcolumntype{C}[1]{>{\centering\let\newline\\\arraybackslash\hspace{0pt}}m{#1}}
\newcolumntype{R}[1]{>{\raggedleft\let\newline\\\arraybackslash\hspace{0pt}}m{#1}}
\newcolumntype{N}{@{}m{0pt}@{}}
\newcommand{\norm}[1]{\left\lVert#1\right\rVert}

\newenvironment{smalleralign}[1][\small]
 {\par\nopagebreak\leavevmode\vspace*{-\baselineskip}%
  \skip0=\abovedisplayskip
  #1%
  \def\maketag@@@##1{\hbox{\m@th\normalfont\normalsize##1}}%
  \abovedisplayskip=\skip0
  \align}
 {\endalign\ignorespacesafterend}

\allowdisplaybreaks

\usepackage{stackengine}
\setstackEOL{\\}
\setstackgap{L}{\normalbaselineskip}

\SetBgContents{\normalsize{%
    \rotatebox{0}{\Longstack{%
    \\© 2020 IEEE. Personal use of this material is permitted. Permission from IEEE must be obtained for all other uses, in any current or future media, including reprinting/\\republishing this material for advertising or promotional purposes, creating new collective works, for resale or redistribution to servers or lists, or reuse of any copyrighted\\ component of this work in other works. This article has been accepted for publication in a future issue of this journal, but has not been fully edited.\\ Content may change prior to final publication. Citation information: https://doi.org/10.1109/LWC.2020.3007417}}}}
\SetBgScale{0.75}
\SetBgOpacity{1}
\SetBgColor{blue}
\SetBgVshift{0.25cm}

\begin{document}
\title{Fundamental Performance Limits of mm-Wave Cooperative Localization in Linear Topologies 
\author{Varun~Amar~Reddy,~\IEEEmembership{Student Member,~IEEE,}~Ahmad Bazzi, ~Gordon~L.~St{\"u}ber,~\IEEEmembership{Fellow,~IEEE,}\\~Suhail~Al-Dharrab,~\IEEEmembership{Member,~IEEE,}
~Wessam~Mesbah,~Ali~Hussein~Muqaibel,~\IEEEmembership{Senior Member,~IEEE,}
\thanks{This work is supported by the Center for Energy and Geo Processing at Georgia Institute of Technology and King Fahd University of Petroleum and Minerals (KFUPM), under research grant number GTEC1601.}
\thanks{V. A. Reddy and G. L. St{\"u}ber are with the School of Electrical and Computer Engineering, Georgia Institute of Technology, Atlanta, GA 30332, USA (e-mail: varun.reddy@gatech.edu; stuber@ece.gatech.edu).}
\thanks{A. Bazzi is with CEVA, Les Bureaux Green Side 5, Bat 6, 400 Avenue Roumanille, 06410 Biot, France (e-mail: ahmad.bazzi@ceva-dsp.com).}%
\thanks{S.~Al-Dharrab, W.~Mesbah, and A.~H.~Muqaibel are with the Electrical Engineering Department, King Fahd University of Petroleum and Minerals, Dhahran 31261, Saudi Arabia (e-mail: suhaild@kfupm.edu.sa; mesbahw@gmail.com; muqaibel@kfupm.edu.sa).}
}}

\markboth{ACCEPTED BY IEEE WIRELESS COMMUNICATIONS LETTERS}{Reddy \MakeLowercase{\textit{et al.}}: Fundamental Performance Limits of mm-Wave Cooperative Localization in Linear Topologies}

\maketitle
\begin{abstract}
In applications such as seismic acquisition, the position information of sensor nodes, that are deployed in a linear topology, is desired with sub-meter accuracy in the presence of a limited number of anchor nodes. This can be achieved with antenna arrays via mm-wave cooperative localization, whose performance limits are derived in this letter. The number of anchor nodes is seen to have a stronger impact than the number of antenna elements in the anchor nodes. Succinct closed-form expressions for the position error bound are also obtained for 1-hop and 2-hop cooperative localization, where sub-meter accuracy is perceived over several hundred nodes. 
\end{abstract}

\section{Introduction}
In seismic acquisition for oil and gas exploration, thousands of sensor nodes, called \textit{geophones}, are deployed linearly along \textit{Receiver Lines (RLs)} across areas of up to 100~km$^{2}$~\cite{ICC,TWC}. Seismic waves are recorded by the geophones to generate an image of the subsurface layers of the Earth. Location information of the geophones is of primary importance, since inaccurate positioning can lead to a degradation of the image quality. Hence, an accuracy of at least 1~m is typically desired~\cite{Savazzi}. Although global positioning system (GPS) modules can provide this level of accuracy, their positioning capability may be unreliable in certain regions. By incorporating localization schemes into the communication system, the need for several thousands of GPS modules can altogether be eliminated, thereby reducing the overall cost.  \par 
The subject of cooperative localization has been comprehensively analyzed in~\cite{Shen1,Shen,Win}. Additional generalizations have been provided in~\cite{Win} for the use of antenna arrays. A Cram\'er-Rao bound (CRB) analysis on the location accuracy has been studied in~\cite{Larsson} for sensor networks based on time-of-arrival (TOA) measurements. With regards to a linear geophone network, the CRB on the positioning accuracy in ultra-wideband (UWB) systems has been analyzed in \cite{Nicoli}. \par 
Recently, there has been a shift towards high-density seismic acquisition, leading to Gigabit rate requirements for data transfer. Hence, mm-wave communication with the IEEE 802.11ad standard has been proposed for real-time data delivery~\cite{VTC18}. Such a network architecture can be augmented with a mm-wave cooperative localization scheme. This letter extends the model in~\cite{mmWave2} for single-anchor localization using classical beamforming with antenna arrays to the cooperative scenario, followed by a generalized derivation of the fundamental performance limits. With high path loss mm-wave propagation, cooperation between nodes typically will not exceed two hops. Closed-form analytical expressions are derived for the CRB and the worst-case position error bound (PEB) in the case of 1-hop and 2-hop cooperation in linear topologies. Such expressions can prove to be computationally advantageous in topologies comprising a large number of nodes.\BgThispage
\section{System Model}
\begin{figure}[!t]
\centering
\includegraphics[width=\linewidth]{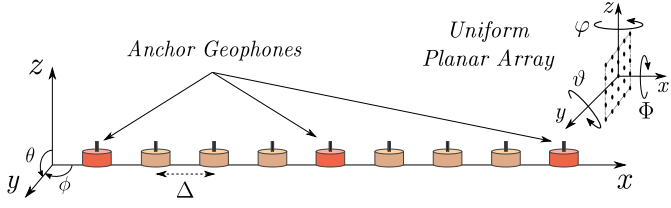}
\caption{A linear topology of geophones along the $x$-axis with $G=6,~W=3,$ and $N_{g}=25$.}
\label{fig:topo}
\end{figure}

Consider an RL comprising $G+W$ geophones along the $x$-axis, with the $g^{th}$ geophone having an unknown location $\mathbf{p}_{g} = [x_{g}, y_{g}, z_{g}]^\text{T} \in \mathbb{R}^{3}$, $1 \leq g \leq G$. There are $W$ anchor geophones with known locations $\mathbf{p}_{g},G+1 \leq g \leq G+W$, that are uniformly positioned along the RL, as shown in Fig.~\ref{fig:topo}. Each geophone is equipped with a uniform planar antenna array of known orientation, having $N_{g}$ number of antenna elements. The orientation of the antenna array at the $g^{th}$ geophone is modelled by the rotation matrices $\mathbf{R}_{z}(\varphi_{g})$ (counter-clockwise rotation about the $z$-axis), $\mathbf{R}_{y}(\vartheta_{g})$ (clockwise rotation about the $y$-axis), and $\mathbf{R}_{x}(\Phi_{g})$ (clockwise rotation about the $x$-axis). As shown in Fig.~\ref{fig:topo}, given that the array is initially aligned vertically and lying in the $yz$-plane, and the position of the $i^{th}$ antenna element is denoted by $\mathbf{p}_{g}^{(i')}$, then its position $\mathbf{p}_{g}^{(i)}$ after rotation is given by $\mathbf{R}_{z}(\varphi_{g})\mathbf{R}_{y}(\vartheta_{g})\mathbf{R}_{x}(\Phi_{g})\mathbf{p}_{g}^{(i')}$. The centroid of the array is located at $\mathbf{p}_{g}$. \par 
Pairwise measurements are performed between all geophones, as per the maximum transmission range (denoted by $R_{max}$), and subsequently transferred to a central processing node. As employed in \cite{mmWave2}, a direct localization scheme is applied at the central node where the position information is estimated directly from all the received signals~\cite{dpd2} with parameters such as the array response and the TOA being implicitly used. Estimation of the orientation of the antenna arrays is not considered, as this information is irrelevant to seismic acquisition. \par 
Let the transmit signal from an antenna element in the $g^{th}$ geophone be denoted as $s_{g}(t)$. 
\begin{smalleralign}[\normalsize]
s_{g}(t) = \sqrt{E/N_{g}} ~\Re \{ p(t) e^{j 2 \pi f_{c} t} \}
\end{smalleralign}
where $E$ is the total transmit energy, $N_{g}$ is the number of antenna elements, $f_{c}$ is the carrier frequency, and $p(t)$ is a unitary pulse-shaping signal whose corresponding Fourier transform is denoted by $P(f)$. In this model, it is assumed that all geophones have coarsely identified virtual antenna sectors by performing pairwise beamforming protocols. For instance, this is achieved through the sector-level sweep in the IEEE 802.11ad standard. Given that the signal bandwidth $B \ll f_{c}$, classical beamforming can be achieved through phased arrays and the transmit beamforming vector $\mathbf{f}_{g}$ at the $g^{th}$ geophone is expressed as 
\begin{smalleralign}[\small]
\mathbf{f}_{g} &= [ \omega_{1}~\omega_{2}~\cdots ~\omega_{N_{g}} ]^\text{T} 
\end{smalleralign}
\begin{smalleralign}[\small]
\omega_{i} &= \exp \left\{ j 2 \pi f_{c} \tau_{g}^{(i)}(\boldsymbol{\theta}_{g}^{(s)}) \right\} &,~~~~ \tau_{g}^{(i)}(\boldsymbol{\theta}_{g}^{(s)}) = \dfrac{\mathbf{d}(\boldsymbol{\theta}_{g}^{(s)}) \mathbf{p}_{g}^{(i)}}{c}
\end{smalleralign}
Considering spherical coordinates, define $\boldsymbol{\theta}=[\theta$,$\phi]^\text{T}$ and $\mathbf{d}(\boldsymbol{\theta})$=$[\sin(\theta)$$\cos(\phi)$,$\sin(\theta)$$\sin(\phi)$,$\cos(\theta)]$ to be the direction cosine vector. The intended steering angle is denoted by $\boldsymbol{\theta}_{g}^{(s)}$. \par 
Let the vector of signals (in the frequency domain) received by all the antenna elements in the $h^{th}$ geophone, corresponding to the transmitted signals from the $g^{th}$ geophone, be denoted as $\mathbf{r}_{h,g}(f) \in \mathbb{C}^{N_{h} \times 1}$.
\begin{smalleralign}[\small]
\mathbf{r}_{h,g}(f) &= \mathbf{x}_{h,g}(f) + \mathbf{n}_{h,g}(f) \label{eq:r(f)} \\
\mathbf{x}_{h,g}(f) &= \sqrt{E/N_\text{g}}~ P(f) ~\mathbf{a}_{h}^{r}(f) \mathbf{a}_{h,g}(f) \mathbf{A}_{g}^{t}(f) \mathbf{f}_{g} \label{eq:x(f)} 
\end{smalleralign}
where $\mathbf{x}_{h,g}(f)$ denotes the vector of the desired received signals and $\mathbf{n}_{h,g}(f)$ is a noise vector with $n_{h,g}(t)$ being modelled as a circularly-symmetric Gaussian random variable with zero mean and variance $N_{0}$, to capture thermal background effects in the receiver. The terms $\mathbf{a}_{h}^{r}(f)$ and $\mathbf{A}_{g}^{t}(f)$ denote the receive and transmit array frequency response matrices respectively, and $\mathbf{a}_{h,g}(f)$ is the channel frequency response vector.
\begin{smalleralign}[\normalsize]
\mathbf{a}_{h}^{r}(f) &= [ \exp(j 2 \pi (f+f_{c}) \tau_{h}^{(m)}(\boldsymbol{\theta}_{h})) ]^\text{T}_{m=1,2,...,N_{h}} \label{eq:rx-array-resp} \\
\mathbf{A}_{g}^{t}(f) &= \text{diag} \{ \exp(-j 2 \pi (f+f_{c}) \tau_{g}^{(i)}(\boldsymbol{\theta}_{h})) \}_{i=1,2,...,N_{g}} \\
\mathbf{a}_{h,g}(f) &= a_{h,g} \exp ( -j 2 \pi (f+f_{c}) \tau_{h,g} ) \mathbf{I}_{1 \times N_{g}} \label{eq:channel-resp} \\
\tau_{h,g} &= \norm { \mathbf{p}_{h} - \mathbf{p}_{g} } /c \label{eq:prop-delay}
\end{smalleralign}
where $\boldsymbol{\theta}_{h}$ is the angle-of-arrival (AOA) with respect to the centroids of both the arrays, and $\mathbf{I}_{1 \times N_{g}}$ is the vector of all ones with dimension $N_{g}$. The terms $a_{h,g}$ and $\tau_{h,g}$ represent the amplitude and the propagation delay of the channel between the $h^{th}$ and $g^{th}$ geophones respectively. Given that $\Delta \gg \mathbb{A}_{g},~\forall g$, where $\mathbb{A}_{g}$ is the antenna aperture of the $g^{th}$ geophone, $a_{h,g} \approx a$~\cite{mmWave2}.
\section{Position Error Bound}
In this section, an expression for the position error bound of the $g^{th}$ geophone is derived. Let $\boldsymbol{\psi} = [ \mathbf{p}^\text{T}_{1} ~ \mathbf{p}^\text{T}_{2} ~ \cdots ~ \mathbf{p}^\text{T}_{G}~ a ]^\text{T} $ be defined as the vector of unknown parameters in the network. The indicator function $I(g,h)$ represents the feasibility of a link between the $g^{th}$ and $h^{th}$ geophones. 
\begin{smalleralign}[\small]
I(g,h) = \begin{cases}
~1 &,~ \vert g-h \vert \Delta \leq R_{max} \\
~0 &,~ \vert g-h\vert \Delta > R_{max}
\end{cases}
\end{smalleralign}
where $\Delta$ is the distance between adjacent geophones. \par
For all $g,h$, let $\mathbf{r}_{h,g}(f)$ and $\mathbf{x}_{h,g}(f)$ be grouped into the vectors $\mathbf{r}(f)$ and $\mathbf{x}(f)$ respectively. The performance of any unbiased estimator $\hat{\boldsymbol{\psi}} = \hat{\boldsymbol{\psi}} (\mathbf{r}(f))$ is bounded by the Cram\'er-Rao bound (CRB)~\cite{Kay}.
\begin{smalleralign}[\small]
\mathbb{E}_{\mathbf{r}|\boldsymbol{\psi}} & \left\{ \left[ \hat{\boldsymbol{\psi}} - \boldsymbol{\psi} \right] \left[ \hat{\boldsymbol{\psi}} - \boldsymbol{\psi} \right]^\text{T} \right \} \succeq \mathbf{J}_{\boldsymbol{\psi}}^{-1} = \text{CRB}(\boldsymbol{\psi}) \\
\mathbf{J}_{\boldsymbol{\psi}} &\triangleq - \mathbb{E}_{\mathbf{r}|\boldsymbol{\psi}} \left\{ \boldsymbol{\nabla}^{2}_{\boldsymbol{\psi} \boldsymbol{\psi}} ~\text{ln}~ f(\textbf{r}|\boldsymbol{\psi}) \right\} \label{eq:FIM}
\end{smalleralign}
where $\mathbf{J}_{\boldsymbol{\psi}}$ is the Fisher Information Matrix (FIM) and $\boldsymbol{\nabla}_{\boldsymbol{\psi}} = \partial/\partial \boldsymbol{\psi}$. The following analysis is undertaken for the case where a-priori position knowledge is unavailable. Considering all the measurements to be statistically independent from one another, the log-likelihood function, $\ln f(\textbf{r} | \boldsymbol{\psi})$, may be rewritten in terms of the function $\sigma(\bullet)$ that serves as a shorthand for the summations in \eqref{eq:decomposed}, where $\alpha_{g,h} \triangleq I(g,h)~\ln f(\textbf{r}_{h,g} | \boldsymbol{\psi})$ and
\begin{smalleralign}[\normalsize]
\text{ln} ~ f(\textbf{r} | \boldsymbol{\psi}) &= ~\sigma(\alpha_{g,h}) \\ &  \hspace{-5mm} = \underbrace{ \sum\limits_{g=1}^{G} ~ \sum\limits_{\substack{h=1 \\ h \neq g}}^{G+W} (\alpha_{g,h}) }_{\substack{\text{Measurements arising from} \\ \text{geophone transmissions}}} + \underbrace{ \sum\limits_{g=G+1}^{G+W} ~ \sum\limits_{\substack{h=1}}^{G} ~(\alpha_{g,h}) }_{\substack{\text{Measurements arising from} \\ \text{anchor geophone transmissions}}} \label{eq:decomposed}
\end{smalleralign}
where $\ln f(\textbf{r}_{h,g} | \boldsymbol{\psi})$ is the log-likelihood ratio of $\mathbf{r}_{h,g}$ as obtained from the Karhunen-Lo\`eve expansion of $\mathbf{r}_{h,g}(f)$ conditioned on $\boldsymbol{\psi}$~\cite{Shen}.
\begin{smalleralign}[\small]
\ln f(\textbf{r}_{h,g}|\boldsymbol{\psi}) &= \dfrac{2}{N_{0}} \int_{B} \mathbf{x}_{h,g}^\text{H}(f)\mathbf{r}_{h,g}(f) df -  \dfrac{1}{N_{0}} \int_{B} \norm{\mathbf{x}_{h,g}(f)}^{2} df
\end{smalleralign}
The FIM $\mathbf{J}_{\boldsymbol{\psi}}$ in \eqref{eq:FIM} can be expressed as \vspace{1mm}
\begin{smalleralign}[\small]
\mathbf{J}_{\boldsymbol{\psi}} &= \left(  {\begin{array}{cccc}
\mathbf{J}_{\boldsymbol{\psi}_{1}, \boldsymbol{\psi}_{1}} & \cdots & \mathbf{J}_{\boldsymbol{\psi}_{1}, \boldsymbol{\psi}_{G}} & \mathbf{J}_{\boldsymbol{\psi}_{1}, a} \\[3pt]
\vdots & \mathbf{J}_{\boldsymbol{\psi}_{u},\boldsymbol{\psi}_{v}} & \vdots & \vdots \\[3pt]
\mathbf{J}_{\boldsymbol{\psi}_{G}, \boldsymbol{\psi}_{1}} & \cdots & \mathbf{J}_{\boldsymbol{\psi}_{G}, \boldsymbol{\psi}_{G}} & \mathbf{J}_{\boldsymbol{\psi}_{G}, a} \\[3pt]
\mathbf{J}_{a, \boldsymbol{\psi}_{1}} & \cdots & \mathbf{J}_{a, \boldsymbol{\psi}_{G}} & \mathcal{J}_{aa}
\end{array} } \right)
\end{smalleralign}
where the notation $\mathbf{J}_{\boldsymbol{\psi}_{u},\boldsymbol{\psi}_{v}} ~( \boldsymbol{\psi}_{u}$ being the $u^{th}$ element of $\boldsymbol{\psi},~u,v = 1,2,...,G+1)$ is defined as
\begin{smalleralign}[\small]
\mathbf{J}_{\boldsymbol{\psi}_{u}, \boldsymbol{\psi}_{v}} & \triangleq - \mathbb{E}_{\mathbf{r}|\boldsymbol{\psi}} \left\{ \boldsymbol{\nabla}^{2}_{\boldsymbol{\psi}_{u} \boldsymbol{\psi}_{v}} ~\text{ln}~ f(\mathbf{r}|\boldsymbol{\psi}) \right\} \label{eq:FIM-element}
\end{smalleralign}
For a given geophone, the measurements at each of the receiving antenna elements are independent~\cite{mmWave2}. Substituting for $\mathbf{r}_{h,g}(f)$ in \eqref{eq:decomposed} from \eqref{eq:r(f)}, and considering $\mathbb{E}_{\mathbf{r}} \{ \mathbf{n}_{h,g}(f) \} = 0 ~\forall g,h$, the expression in \eqref{eq:FIM-element} can be expanded as
\begin{smalleralign}[\small]
\mathbf{J}_{\boldsymbol{\psi}_{u}, \boldsymbol{\psi}_{v}} &= \sigma \left( \mathbf{J}^{(h,g)}_{\boldsymbol{\psi}_{u},\boldsymbol{\psi}_{v}} \right) \label{Jpsi-coop} \\
\mathbf{J}^{(h,g)}_{\boldsymbol{\psi}_{u},\boldsymbol{\psi}_{v}} &= \dfrac{2~I(g,h)}{N_{0}} \sum\limits_{m=1}^{N_{h}} \int_{B} \Re \left\{ \dfrac{\partial X_{h,g,m}^{*}(f)}{\partial \boldsymbol{\psi}_{u}}  \dfrac{\partial X_{h,g,m}(f)}{\partial \boldsymbol{\psi}_{v}} \right\} df \label{eq:J_hg}
\end{smalleralign}
where $X_{h,g,m}(f)$ is the $m^{th}$ element of the vector $\mathbf{x}_{h,g}(f)$. 
Following the analysis technique in~\cite{Shen,Win},
\begin{smalleralign}[\small]
\mathbf{J}_{\boldsymbol{\psi}_{u},\boldsymbol{\psi}_{v}} &= \begin{cases}
\sum\limits_{\substack{g=1 \\ g \neq u}}^{G+W} \left( \mathbf{J}^{(u,g)}_{\boldsymbol{\psi}_{u},\boldsymbol{\psi}_{u}} + \mathbf{J}^{(g,u)}_{\boldsymbol{\psi}_{u},\boldsymbol{\psi}_{u}} \right) &, u=v; 1 \leq u,v \leq G \vspace*{2mm}\\
-
\left( \mathbf{J}^{(u,v)}_{\boldsymbol{\psi}_{u},\boldsymbol{\psi}_{u}} + \mathbf{J}^{(v,u)}_{\boldsymbol{\psi}_{u},\boldsymbol{\psi}_{u}} \right) &, u \neq v; 1 \leq u,v \leq G \vspace*{2mm}\\
0 &, 1 \leq u \leq G, v=G+1 \vspace*{2mm}\\
0 &, 1 \leq v \leq G, u=G+1 \vspace*{2mm}\\
\mathcal{J}_{aa} &, u=v=G+1
\end{cases}  \label{eq:Jpsi_uv}
\end{smalleralign}
The entries of the matrix $\mathbf{J}_{\boldsymbol{\psi}_{u},\boldsymbol{\psi}_{v}}$ are derived in Appendix A, where it is seen that rotation about the $x$-axis does not impact the nature of the FIM for a topology along the $x$-axis, an observation that was also made in~\cite{mmWave2}. Additionally, channel reciprocity does not itself imply equivalent bidirectional measurements between any two geophones, since the array orientation and the number of antenna elements at the transmit and receive sides can differ. The PEB of the $g^{th}$ geophone, denoted by $\text{PEB}_{g}$, can be computed from the $g^{th}$ $(3 \times 3)$ submatrix occuring along the diagonal of $\text{CRB}(\boldsymbol{\psi})$.
\begin{equation}
\text{PEB}_{g} = \sqrt{\text{tr} \left\{ [ \text{CRB}(\boldsymbol{\psi}) ]_{\scaleto{g,g}{4pt}}\right\} } = \sqrt{\text{tr} \{ [ \mathbf{J}_{\boldsymbol{\psi}}^{-1} ]_{\scaleto{g,g}{4pt}} \} } \label{eq:peb}
\end{equation}
\section{Special Cases}
In this section, closed-form analytical expressions for the maximum position error bound are derived for a linear topology in two specific scenarios: $R = \Delta$ and $R = 2 \Delta$. Consider a total of $G$ geophones with $W=2$, and $N_{g} = N, \vartheta_{g} = \vartheta', \varphi_{g} = \varphi',~\forall g$, such that symmetric measurements are obtained between any pair of geophones. In particular, the PEB of the $(G/2)^{th}$ geophone lying at the center of the topology is derived, as it perceives the worst-case localization performance. For simplicity, assume that $G$ is an even value\footnote{When $G$ is even, the $(G/2)^{th}$ and $(G/2+1)^{th}$ geophones have equal and maximum position error. When $G$ is odd, the $((G+1)/2)^{th}$ geophone has maximum position error.}.
\subsection{One-hop Cooperation}
\label{section:1hop}
When $R = \Delta$, measurements are made by a geophone only up to one hop away, resulting in a total of two measurements for each geophone. Under these conditions, $\mathbf{J}_{\boldsymbol{\psi}}$ takes the form of a direct sum (denoted by $\oplus$) of a  Toeplitz tridiagonal symmetric block matrix (denoted by tri\{$\bullet,\bullet$\}) and $\mathcal{J}_{aa}$. \vspace*{2mm}
\begin{smalleralign}[\small]
\mathbf{J}_{\boldsymbol{\psi}} &= \text{tri}\{\mathbf{J}_{A},\mathbf{J}_{B} \} \oplus \mathcal{J}_{aa} \\[3pt]
\mathbf{J}_{A} &= 4 \times \text{diag}\{\mathcal{J}_{xx},\mathcal{J}_{yy},\mathcal{J}_{zz}\} &,~~~~~ \mathbf{J}_{B} = -(1/2) \times \mathbf{J}_{A}
\end{smalleralign}
where the FIM elements $\mathcal{J}_{xx},\mathcal{J}_{yy},\mathcal{J}_{zz}$ are the entries of the matrix in \eqref{eq:finalFIM} when $\left( g-h \right) = \pm 1$. Since the inverse of $\text{tri}\{\mathbf{J}_{A},\mathbf{J}_{B} \}$ is sufficient to obtain an expression for the CRB submatrix for the $(G/2)^{th}$ geophone (denoted by $\mathbf{C}_{G/2}$), the mathematical technique in \cite{TriInverse} is utilized to compute the inverse of a tridiagonal Toeplitz block matrix. \vspace*{2mm}
\begin{smalleralign}[\small]
\mathbf{C}_{G/2}  &=  \left( \mathbf{J}_{A} - \mathbf{X}_{G/2} - \mathbf{Y}_{G/2} \right)^{-1} \label{CRB1} \\[5pt]
\mathbf{X}_{G/2}  &=  \left(  {\begin{array}{cc}
\mathbf{0}_{3} & \mathbf{J}_{B} \\
-\mathbf{J}_{B} & \mathbf{J}_{B} \mathbf{J}_{A}
\end{array} } \right)^{G/2} ~\otimes ~ \mathbf{0}_{6} \\[5pt]
\mathbf{Y}_{G/2}  &=  \left(  {\begin{array}{cc}
\mathbf{0}_{3} & \mathbf{J}_{B} \\
-\mathbf{J}_{B} & \mathbf{J}_{B} \mathbf{J}_{A}
\end{array} } \right)^{G/2-1} ~\otimes ~ \mathbf{0}_{6} \label{CRB3}
\end{smalleralign}
where $\otimes$ denotes the matrix M{\"o}bius transformation and $\mathbf{0}_{n}$ denotes the zero matrix of dimension $n \times n$. Solving \eqref{CRB1}-\eqref{CRB3}, a closed-form expression\footnote{In general, the CRB submatrix for the geophone at the center of the topology is given by $\{2(G+1)^2-1+(-1)^{G+1}\}/ \{4(G+1)\} \times \mathbf{J}_{A}^{-1}$.} for $\mathbf{C}_{G/2}$ is obtained as \vspace*{2mm}
\begin{smalleralign}[\small]
\mathbf{C}_{G/2} &= \dfrac{G(G+2)}{2 (G+1)} \times \mathbf{J}_{A}^{-1} \\[4pt]
\text{PEB}_{G/2} &= \sqrt{ \text{tr} \left\{ \mathbf{C}_{G/2} \right\}} \label{eq:1hop-peb}
\end{smalleralign}
Hence, the value of $\text{PEB}_{G/2}$ is essentially given by the single-hop parameters $\mathcal{J}_{xx},\mathcal{J}_{yy},\mathcal{J}_{zz}$ that are scaled by a factor which grows exponentially with $G$. 
\subsection{Two-hop Cooperation}
\label{section:2hop}
When $R = 2\Delta$, measurements are made by a geophone up to two hops away. The FIM $\mathbf{J}_{\boldsymbol{\psi}}$ is obtained as the direct sum of a pentadiagonal matrix and $\mathcal{J}_{aa}$. A first simplification is performed by expressing the pentadiagonal matrix as a product of two Toeplitz tridiagonal matrices~\cite{PentaInverse}. \vspace*{2mm}
\begin{smalleralign}[\small]
\mathbf{J}_{\boldsymbol{\psi}} &= \left\{ (2\mathbf{J}_{B}) \ast (\mathbf{T}_{1} \times \mathbf{T}_{2}) \right\} \oplus \mathcal{J}_{aa} \\[5pt]
\mathbf{T}_{1},\mathbf{T}_{2} &= \text{tri}\{(2\mathbf{J}_{B})^{-1}(\mathbf{J}_{A} \pm (\mathbf{J}_{A} + 4 \mathbf{J}_{B})) , \mathbf{I}_{3}\}\\[5pt]
\mathbf{J}_{A} &= -\text{diag}\{\mathcal{J}_{xx},\mathcal{J}_{yy},\mathcal{J}_{zz}\}  , \mathbf{J}_{B} = \text{diag}\{ 4d,d,d \} \times \mathbf{J}_{A} \\[5pt]
d &= L_{0}^{-2\Delta}/(2\Delta)^{4} \times (L_{0}^{-\Delta}/(\Delta)^{4})^{-1} = L_{0}^{-\Delta}/16 
\end{smalleralign}
where $\mathbf{X} \ast \mathbf{Y}$ denotes a blockwise product of $\mathbf{X}$ and each of the sub-blocks of $\mathbf{Y}$, and $d$ denotes the ratio of the FIM element $\mathcal{J}_{yy}$ in the 2-hop case to the 1-hop case, assuming that the path loss arises from free-space propagation and atmospheric absorption. The atmospheric absorption loss is expressed as $L_{0}^{-\Delta}$, where $L_{0} = 10^{0.0017}$ at $f_{c} = 60$~GHz~\cite{VTC18}. \par
An approximation is made by considering only the product of the $(G/2)^{th}$ submatrices of the inverse of each of the tridiagonal matrices, since this product is the dominant contributor to the value of $\mathbf{C}_{G/2}$. \vspace*{2mm}
\begin{smalleralign}[\small]
\text{PEB}_{G/2} & = \sqrt{ \text{tr} \left\{ \mathbf{C}_{G/2} \right\}} \label{eq:penta-peb} \\[4pt]
\mathbf{C}_{G/2} & \approx \left[ \mathbf{T}_{2}^{-1} \right]_{G/2,G/2} \times  \left[ \mathbf{T}_{1}^{-1} \right]_{G/2,G/2} \times (2\mathbf{J}_{B})^{\scaleto{-1}{4pt}} \label{eq:penta-crb1}  \\[4pt]
& = \dfrac{G(G+2)}{2(G+1)} \times (2\mathbf{J}_{B})^{-1} \times  \\[4pt]
&  ~~~~\text{diag}\{p(1/(4d)+2),p(1/d+2),p(1/d+2)\}  \\[4pt]
p(d_{1}) &= \left( \dfrac{r_{1}^{2}-r_{2}^{2}}{s_{1}r_{3}} \right) \left( \dfrac{s_{1}^{2}}{s_{2}} \right)^{G/2-1}  \\[4pt]
s_{1} &= \dfrac{1}{2} \left(d_{1}+\sqrt{d_{1}^{2}-4} \right) \hspace*{5mm},\hspace*{5mm}  s_{2} = d_{1} s_{1}-1  \\[4pt]
r_{2},r_{1} &= \pm \dfrac{\left(d_{1}-1\right)\,\left(4\,d_{1}-(d_{1}^2-2)\,\sqrt{d_{1}^{2}-4}-d_{1}^3\right)}{4\,(d_{1}^{2}-4)\,d_{1}}  \\[4pt]
r_{3} &= \dfrac{1}{2} \left( d_{1}^2-\frac{3\,d_{1}-d_{1}^3}{\sqrt{d_{1}^{2}-4}}-1 \right)
\end{smalleralign}
The above expressions reveal that the value of $\text{PEB}_{G/2}$ grows exponentially with $G$, as observed in the case of 1-hop cooperation. However, the rate of growth is decreased as a function of $d$, thereby improving the overall localization performance.
\section{Performance Evaluation}
Consider a linear topology of geophones, operated by the IEEE 802.11ad standard, with $\Delta=25$~m along the $x$-axis. As per the standard, $f_{c} = 60$~GHz, $B = 2.16$~GHz, and $p(t)$ is a root raised-cosine pulse with roll-off factor = 0.6. The transmit power, noise figure, and system temperature are considered to be 20~dBm, 4~dB, and 300~K respectively. \par
\begin{figure*}[t!]
    \centering
    \subfloat[Variation of $\text{PEB}_{g}$, with all geophones having 25 antenna elements.]{\begin{minipage}[t!]{0.67\columnwidth}
        \centering
        \includegraphics[width=\linewidth]{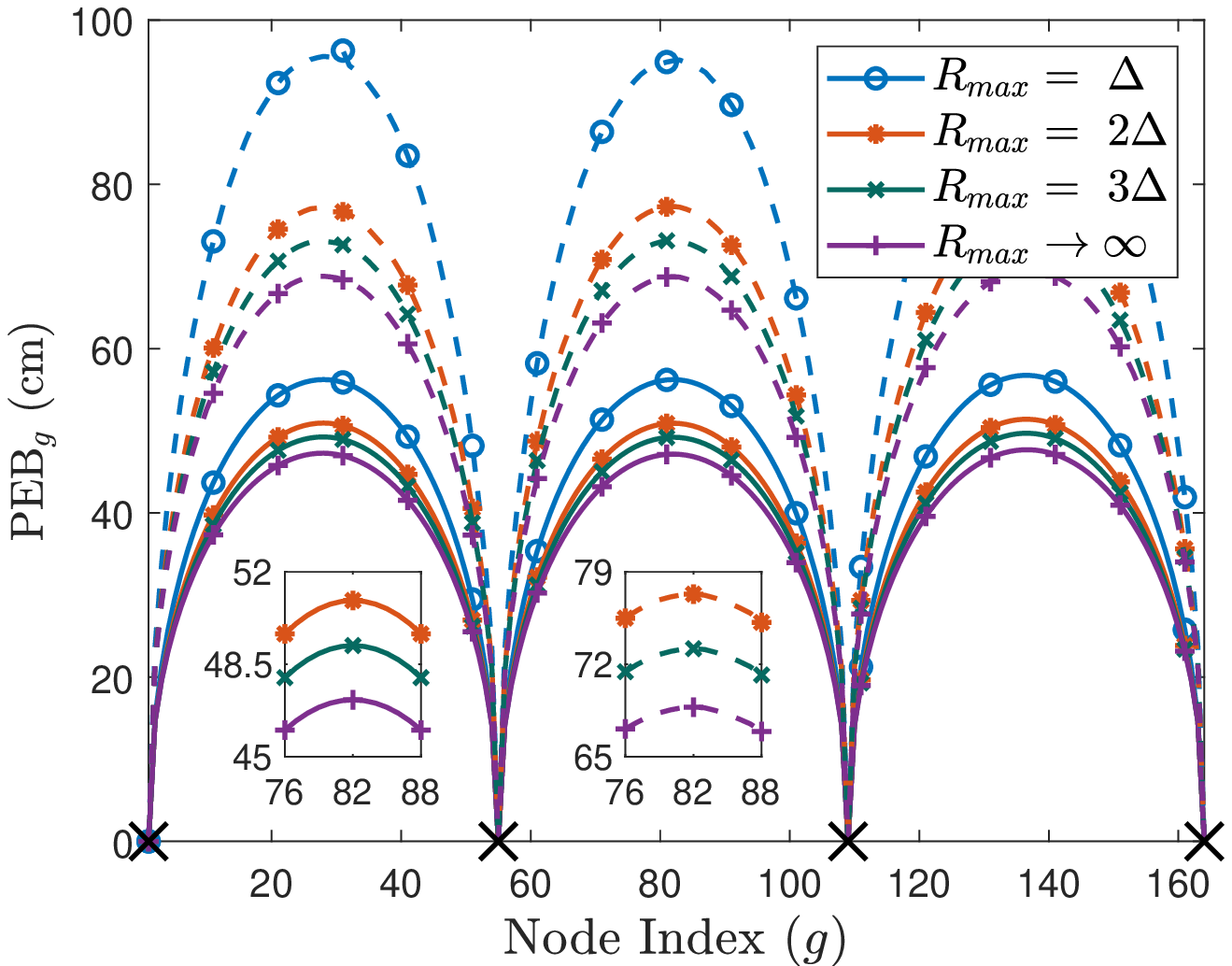}
       \label{fig:peb_4_4}
    \end{minipage}}
    \hfill
    \subfloat[Variation of $\text{PEB}_{g}$, with the anchor geophones having 100 antenna elements.]{\begin{minipage}[t!]{0.67\columnwidth}
        \centering
        \includegraphics[width=\linewidth]{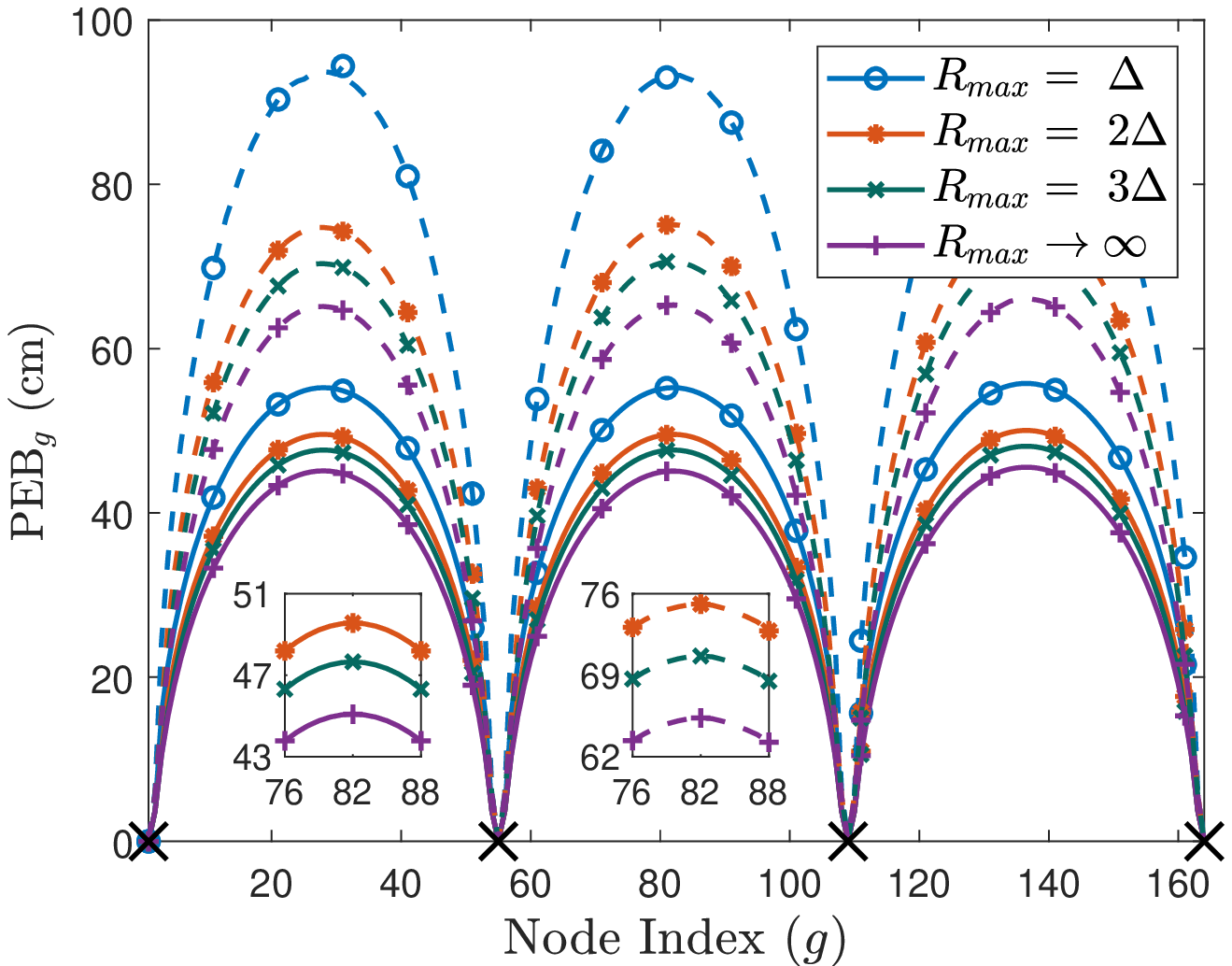}
       \label{fig:peb_4_100}
    \end{minipage}}
	\hfill
    \subfloat[Performance comparison between this work and~\cite{Nicoli} (1-D bound).]{\begin{minipage}[t!]{0.67\columnwidth}
        \centering
        \includegraphics[width=\linewidth]{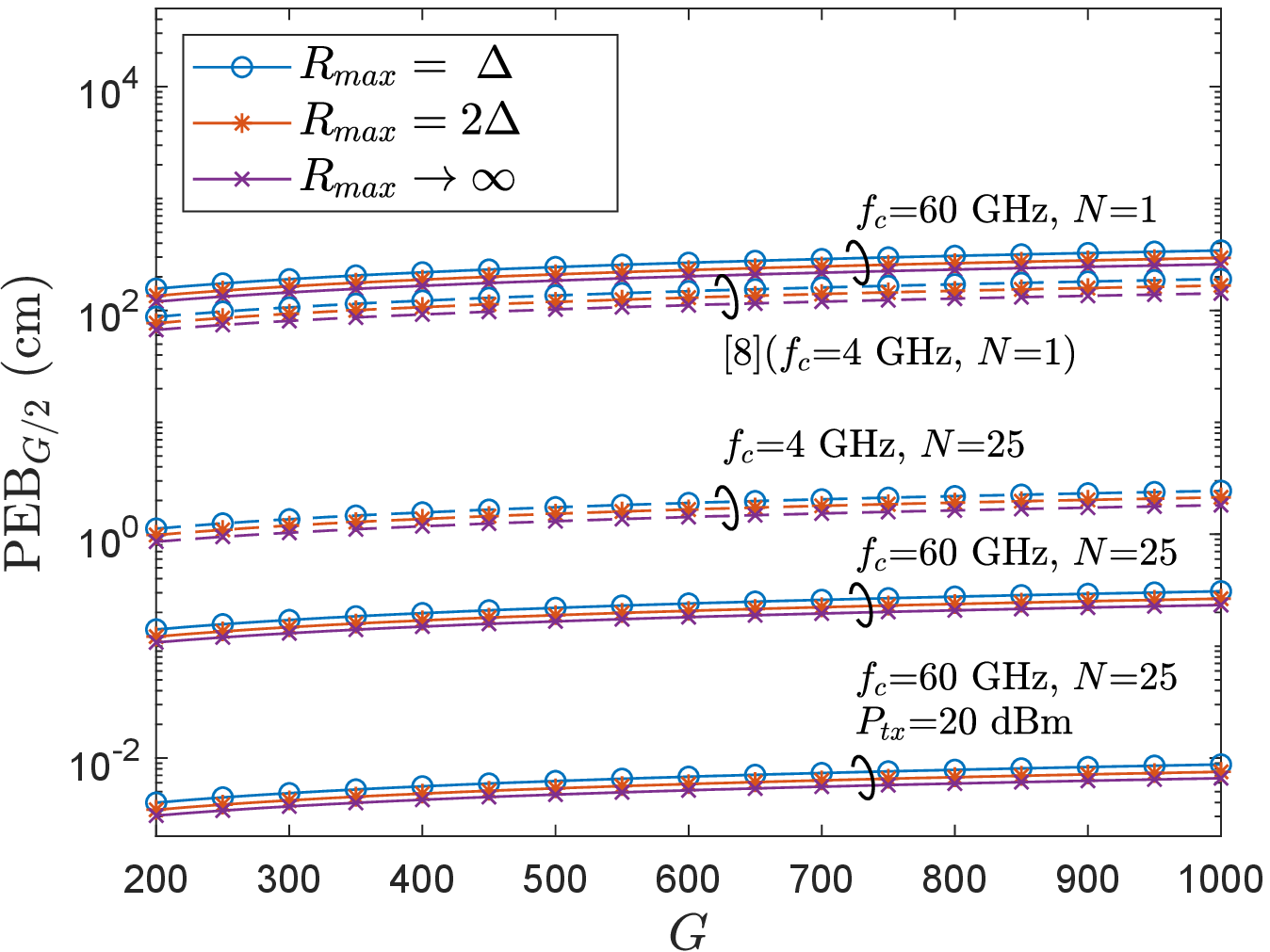}
       \label{fig:UWB_compare_x}
           \end{minipage}}
           \\[1pt]
    \subfloat[The maximum total number of geophones $G$ that ensures a sub-meter value for $\text{PEB}_{G/2}$ as a function of $W$, $N$, and the array orientation.]{\begin{minipage}[t!]{0.6\columnwidth}
    \vspace*{-0.5cm}
        \centering
	\resizebox{\linewidth}{!}{%
	\setlength\extrarowheight{2.5pt}
\normalsize
    \begin{tabular}{|C{0.1\columnwidth} |C{0.225\columnwidth} | C{0.225\columnwidth} | C{0.225\columnwidth} | C{0.225\columnwidth} |}
\hline
\multirow{3}{*}{$W$} & \multicolumn{4}{C{0.9\columnwidth}|}{\textbf{Vertical Orientation}} \\ \cline{2-5}
 & \multicolumn{2}{C{0.45\columnwidth}|}{$R_{max}=\Delta$} & \multicolumn{2}{C{0.45\columnwidth}|}{$R_{max}=2\Delta$} \\ \cline{2-5}
 & $N=25$ & $N=36$  & $N=25$  & $N=36$  \\ \hline
2 & 160 & 475 & 170 & 510 \\ \hline
4 & 470 & 1435 & 500 & 1520  \\ \hline
\multicolumn{5}{C{\columnwidth}}{\vspace{-0.5cm}}\\
\hline
\multirow{3}{*}{$W$} & \multicolumn{4}{C{0.9\columnwidth}|}{\textbf{Averaged Orientation}} \\ \cline{2-5}
 & \multicolumn{2}{C{0.45\columnwidth}|}{$R_{max}=\Delta$} & \multicolumn{2}{C{0.45\columnwidth}|}{$R_{max}=2\Delta$} \\ \cline{2-5}
 & $N=25$ & $N=36$  & $N=25$  & $N=36$  \\ \hline
2 & 50 & 135 & 65 & 190  \\ \hline
4 & 115 & 315 & 160 & 515  \\ \hline
\end{tabular}}
\vspace*{0.5cm}
       \label{tab:listG}
\end{minipage}}
    \hfill
    \subfloat[The maximum value of $G$ that can be accommodated in the case of 1-hop cooperation.]{\begin{minipage}[t!]{0.67\columnwidth}
        \centering
        \includegraphics[width=\linewidth]{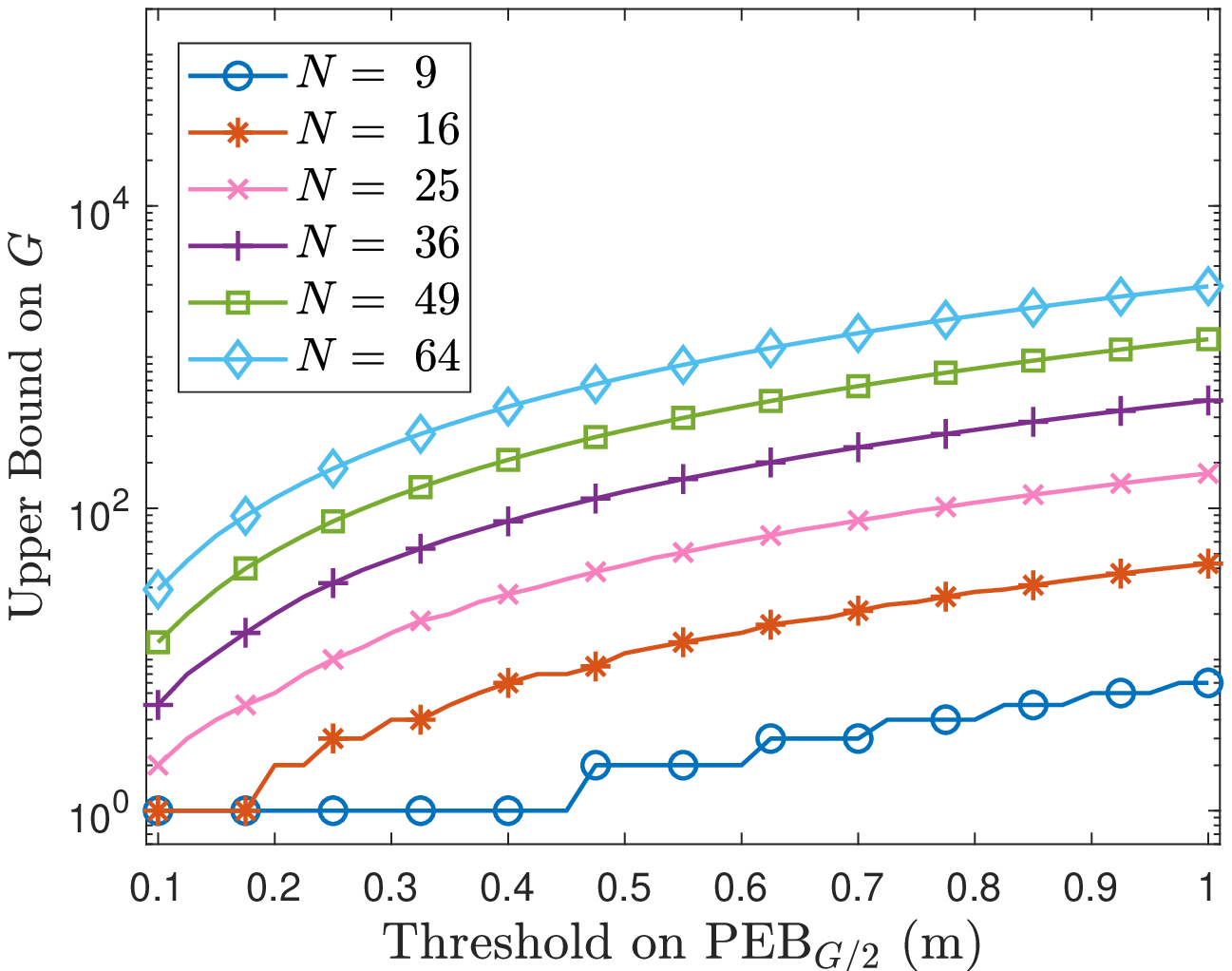}
       \label{fig:1hop-2}
    \end{minipage}}
    \hfill
    \subfloat[Comparison between the true and approximate values of $\text{PEB}_{G/2}$ in the case of two-hop cooperation.]{\begin{minipage}[t!]{0.67\columnwidth}
        \centering
        \includegraphics[width=\linewidth]{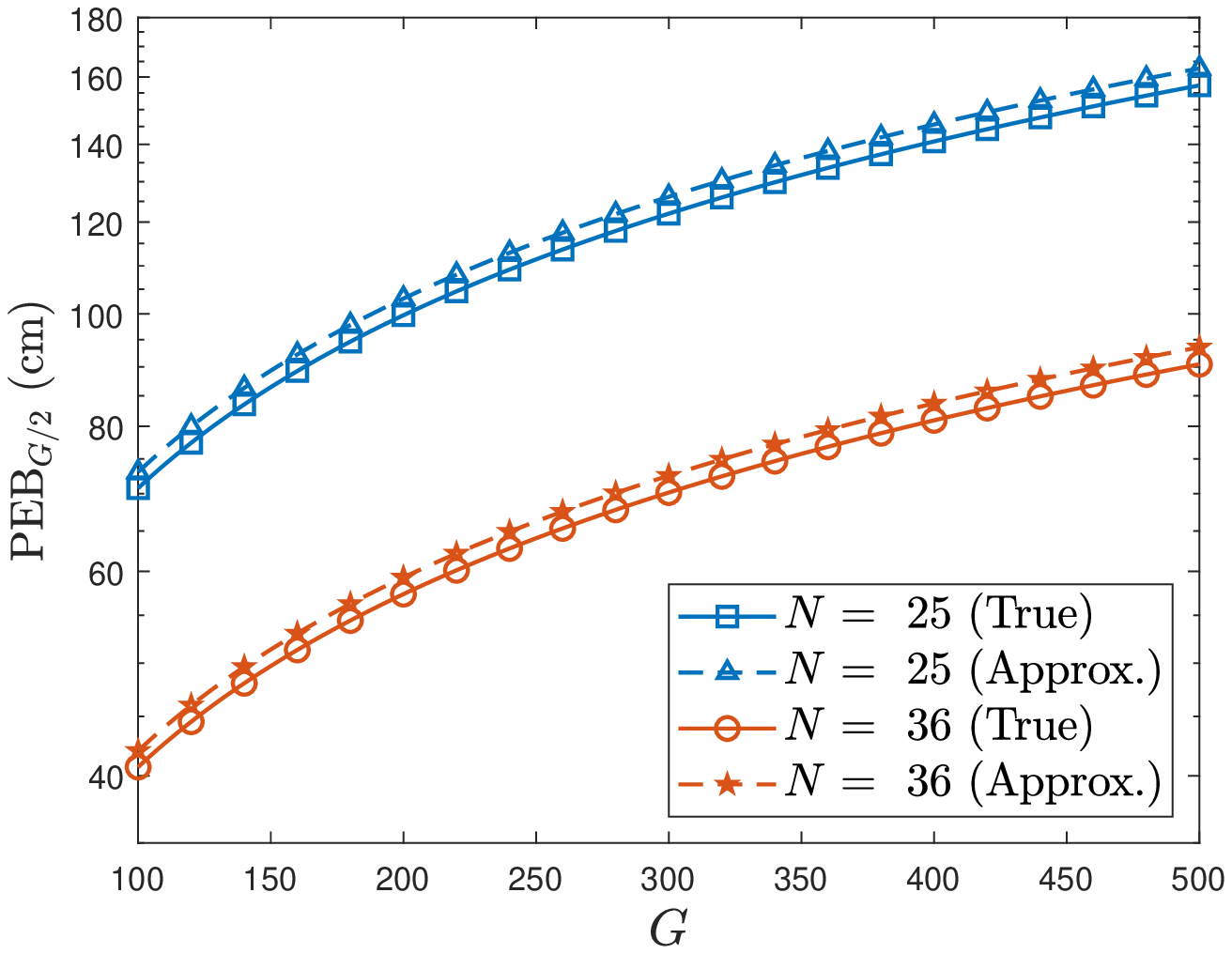}
       \label{fig:2hop}
    \end{minipage}}
           \caption{Performance evaluation of the proposed bounds with respect to various parameters such as $G$, $W$, $R_{max}$, $N_{g}$, and the antenna array orientations at the geophones.}
           \vspace*{-6mm}
\end{figure*}

 
The variation of $\text{PEB}_{g}$ as a function of $g$ is shown in Fig.~\ref{fig:peb_4_4} for $G=160$ and $W=4$ under free-space propagation, where the value of $\text{PEB}_{g}$ is averaged over 1000 Monte Carlo trials. In each trial, $\vartheta_{g}$ and $\varphi_{g}$ are randomly drawn from a uniform distribution in the interval $[0,2\pi)$ (marked by dashed lines). The anchor geophones have been marked by an `$\boldsymbol{\times}$'. In comparison, the PEB can be reduced by 35-40\% when vertical orientation is imposed i.e. $\vartheta_{g}=\varphi_{g}=0,~\forall g$ (marked by solid lines). The number of antenna elements in the anchor geophones is increased to 100 in Fig.~\ref{fig:peb_4_100}, where a marginal improvement in the localization performance is observed, suggesting that the number of anchor geophones has a stronger impact than the number of antenna elements in the anchor geophones. It can also be observed that an increase in the value of $R_{max}$ provides diminishing returns on the PEB, due to the fact that the path loss is more acute at larger distances. \par  
With respect to seismic acquisition, a performance comparison between the proposed bounds ($N_{g}=N,~\forall g$) and those given in~\cite{Nicoli} is made for $B=2.16$ GHz and $W=4$. In~\cite{Nicoli}, an ultra-wideband (UWB) TOA-based cooperative localization scheme\footnote{For a given TOA-based range measurement, the position accuracy is bounded by $ c/(2\sqrt{2}\pi\beta\sqrt{\text{SNR}})$, where SNR is the signal-to-noise ratio and $\beta$ is the effective bandwidth~\cite{Shen} defined as $\sqrt{\int_{-\infty}^{\infty} f^{2} \vert P(f) \vert^{2} df / \int_{-\infty}^{\infty} \vert P(f) \vert^{2} df}$.} is employed with single antennas. The value of $f_{c}$ is set to 4~GHz in the case of~\cite{Nicoli} (within the designated spectrum for UWB systems). The value of $P_{tx}$ is set to $-8$~dBm in both cases since UWB systems are limited by a maximum power spectral density of $-41.3$ dBm/MHz. The two-ray propagation model is considered in both scenarios, since low antenna heights of 0.1-0.2~m are typically perceived in geophone networks for logistical purposes and to counter the effect of high wind speeds~\cite{VTC18}. Hence, the PEB is averaged over 1000 Monte Carlo iterations, with the antenna heights and the orientation angles being uniformly distributed in the intervals $[0.1~\text{m}, 0.2~\text{m}]$ and $[0,2\pi)$ respectively. Since the analysis in~\cite{Nicoli} caters to only a 1-D localization problem, a comparison between the bounds on the $x$-coordinate is presented in Fig.~\ref{fig:UWB_compare_x}. \par 
For the single-antenna case ($N$=1), the $\text{PEB}$ is slightly larger when $f_{c}=60$ GHz, as compared to when $f_{c}=4$ GHz, due to a marginally lower SNR for the given range of antenna heights. For fairness of comparison, the system in~\cite{Nicoli} is also analyzed for the use of antenna arrays with $N=25$ and it is observed that the PEB is lower in the case of $f_{c} = 60$ GHz. As compared to the PEB in~\cite{Nicoli}, the bound given by \eqref{eq:peb} is reduced in proportion to $(\sqrt{1+f_{c}^{2}/\beta^{2}})$ $\times$ $(N)$, wherein the first term arises from additional AOA information provided by the antenna array~\cite{Win} and the second term arises from the beamforming gain. Additionally, a much higher value for $P_{tx}$ is permissible in IEEE 802.11ad systems, that can further reduce the PEB. \par  
Considering the two-ray propagation model, the maximum values for $G$ are listed in Fig.~\ref{tab:listG} such that the value of $\text{PEB}_{G/2}$ (for all three coordinates) does not exceed 1~m. As compared to the case with vertically oriented antenna arrays, the maximum value of $G$ that can be supported in the case with averaged orientation is decreased by 70-75\% for $R_{max}=\Delta$, and by 60-65\% for $R_{max}=2\Delta$. Nevertheless, the proposed bounds imply that sub-meter positioning accuracies can be attained in a highly scalable manner. \par 
With reference to the analysis in Section IV, the closed-form expression for $\text{PEB}_{G/2}$ given by \eqref{eq:1hop-peb} is used to quickly compute the upper bound on the value of $G$ in Fig.~\ref{fig:1hop-2} for $\vartheta'=\varphi'=0$. This performance is further enhanced in the scenario of 2-hop cooperation in Fig.~\ref{fig:2hop}, where the approximation formula for $\text{PEB}_{G/2}$ in \eqref{eq:penta-peb} is shown to be very close to the true value given by \eqref{eq:peb}. 
\section{Conclusion}
In this study, fundamental performance limits have been derived for cooperative mm-wave localization using classical beamforming. The impact of the number of antenna elements, the number of anchor nodes, and the array orientation has been analyzed. Motivated by the fact that cooperation beyond two hops is difficult to achieve under the constraints of mm-wave propagation, closed-form expressions for the maximum PEB in one-hop and two-hop cooperation have been derived. With regards to seismic acquisition, the proposed bounds imply high scalability, with just four anchor nodes yielding sub-meter values for the PEB across 450-1500 nodes (spanning 11-37~km) when all the antenna arrays are vertically oriented, and across 100-500 nodes (spanning 2-12~km) when a statistical average is taken over the antenna arrays' orientations.
\section*{Appendix A}
For $1 \leq u,v \leq G$, the matrix $\mathbf{J}^{(h,g)}_{\boldsymbol{\psi}_{u},\boldsymbol{\psi}_{u}} = \mathbf{J}^{(h,g)}_{\boldsymbol{p}_{u},\boldsymbol{p}_{u}}$ in \eqref{eq:Jpsi_uv} is non-zero only when $u=h$ or $u=g$. An expression is derived in \eqref{eq:finalFIM} by applying the analysis technique described in~\cite{mmWave2} to a linear topology of geophones placed along the $x$-axis ($\boldsymbol{\theta}_{g} = [\pi/2,0]^\text{T}, ~g=1,2,...,G+W$), 
\begin{figure*}[b!]
\hrulefill
\begin{smalleralign}[\small]
\mathbf{J}^{(h,g)}_{\mathbf{p}_{h},\mathbf{p}_{h}} &= \dfrac{8 \pi^{2} \gamma^{(h,g)} (\beta^{2}+f_{c}^{2})}{c^{2}} \left( \begin{array}{ccc} N_{g}N_{h} & 0 & 0 \\[3pt] 0 & \dfrac{N_{g} (N_{h}-1) \mathbb{A}_{h} }{12 ((g-h)\Delta)^{2}} \times (\cos^{2}(\varphi_{h}) + \sin^{2}(\varphi_{h})\sin^{2}(\vartheta_{h})) & 0 \\[3pt]
0 & 0 & \dfrac{N_{g} (N_{h}-1) \mathbb{A}_{h} }{12 ((g-h)\Delta)^{2}} \times \cos^{2}(\vartheta_{h}) \end{array} \right) \label{eq:finalFIM}
\end{smalleralign}
\end{figure*}
where $\mathbb{A}_{h} = (N_{h} \lambda^{2}/4)$ and $\gamma^{(h,g)}$ is the perceived signal-to-noise ratio (SNR) between the $h^{th}$ and $g^{th}$ geophones. For $1 \leq u \leq G, v = G+1$, 
\begin{smalleralign}[\small]
\mathbf{J}_{\boldsymbol{\psi}_{u},\boldsymbol{\psi}_{v}} &= 
\sum\limits_{\substack{g=1 \\ g \neq u}}^{G+W}  \mathbf{J}^{(g,u)}_{\boldsymbol{p}_{u},a} + \mathbf{J}^{(u,g)}_{\boldsymbol{p}_{u},a}  = \sum\limits_{\substack{g=1 \\ g \neq u}}^{G+W}  \mathbf{J}^{(g,u)}_{\boldsymbol{p}_{u},a} + (-\mathbf{J}^{(g,u)}_{\boldsymbol{p}_{u},a})  = 0 \nonumber
\end{smalleralign}
\bibliography{IEEEabrv,Ref}
\bibliographystyle{IEEEtran}
\end{document}